\documentclass[epsfig,12pt]{article}
\usepackage{epsfig}
\usepackage{graphicx}

\newcommand{\beq}{\begin{equation}}   
\newcommand{\eeq}{\end{equation}}
\newcommand{\beqn}{\begin{eqnarray}}   
\newcommand{\eeqn}{\end{eqnarray}}

\newcommand{\gsim}{\lower.7ex\hbox{$
\;\stackrel{\textstyle>}{\sim}\;$}}
\newcommand{\lsim}{\lower.7ex\hbox{$
\;\stackrel{\textstyle<}{\sim}\;$}}

\begin{document}

\begin{titlepage}

\begin{flushright}
FTPI-MINN-13/11, UMN-TH-3143/13\\
%November 15/2012
\end{flushright}

\vspace{0.7cm}

\begin{center}
{  \large \bf  Abrikosov-Nielsen-Olesen string with Non-Abelian\\[2mm] Moduli and Spin-Orbit Interaction}
\end{center}
\vspace{0.6cm}

\begin{center}
 {\large 
 M. Shifman,$^a$ and A. Yung$^{a,b}$}

\end {center}

 \vspace{3mm}
 
\begin{center}

$^a${\em William I. Fine Theoretical Physics Institute, University of Minnesota,
Minneapolis, MN 55455, USA} \\[1mm]
$^{b}${\em Petersburg Nuclear Physics Institute, Gatchina, St. Petersburg
188300, Russia
}

\end {center}

\vspace{2cm}

\begin{center}
{\large\bf Abstract}
\end{center}

It is generally believed that the spontaneous breaking of the Poincar\'e group by flux tubes
(strings) generate only two zero modes localized on the string and associated with the spontaneous breaking
of translational invariance (the so-called Low-Manohar argument). Being perfectly true in many instances
this argument is nevertheless nonuniversal, and has to be amended in the case of order parameters carrying 
spatial indices. We show that under certain circumstances
additional zero (or quasizero) modes can appear due to spin symmetry.

\hspace{0.3cm}

\end{titlepage}

\newpage

The Goldstone theorem \cite{1} relating spontaneously broken global symmetries to the Goldstone particles (massless, or gapless  excitations) is one of the important theoretical cornerstones. If the pattern of the symmetry breaking 
is $G\to H$, where $G$ is an internal symmetry group and $H$ is its subgroup  the number
of the Goldstone particles is $\nu_{\rm r} = {\rm dim} G - {\rm dim} H$ in
the relativistic theories (with the dispersion law $E\sim p$) and $\nu_{\rm nr} = ({\rm dim} G - {\rm dim} H)/2$ in
the nonrelativistic theories (with the dispersion law $E\sim p^2$) \cite{2}. This fact is widely used both in high-energy and condensed-matter theories. The above counting is inapplicable, however, to
geometric (space-time) symmetries \cite{3} (and references therein). For instance, four-dimensional scalar electrodynamics with the Higgs potential for the complex scalar field enjoys the full Poincar\'e symmetry, with 10 generators. The Abrikosov-Nielsen-Olesen string \cite{ANO} breaks the Poincar\'e group down to the 
two-dimensional ``Poincar\'e group" (three generators) times U(1) (one generator). Naively
one may expect six zero modes. However, we have only two Goldstone excitations on the string world sheet 
(one in the nonrelativistic case)
due to the
fact that not all broken generators  are  independent operators on the string world sheet \cite{3}. 

In this paper we will show that 
the analysis of the spontaneous breaking of the geometric (space-time) symmetries  \cite{3} contains a subtle point that can invalidate the above common wisdom and lead to the occurrence of extra moduli fields (gapless excitations, Goldstone particles) 
on the string world sheet. Such extra moduli fields can appear in models with the order parameters carrying spatial indices, such as 
those relevant for superfluidity in $^3$He (see e.g. \cite{5}).  This example was studied in the recent publication \cite{6},
which in fact inspired the general consideration presented below.

The necessary and sufficient condition for the extra massless excitations to appear on the string world sheet is
the existence of a limit in which a spacial symmetry of the bulk (e.g. the O(3) rotation symmetry) becomes enhanced and (a part) of the enhanced symmetry acts as an internal symmetry spontaneously broken on the string. In these circumstances the O(3) part of the Poincar\'e group which, being spontaneously broken, is {\em not} represented by 
independent gapless excitations in the standard treatment \cite{3}, acquires its own non-Abelian moduli fields, the Goldstone bosons localized on the string.\footnote{Nonperturbative effects tending to generate a mass gap in two dimensions can be made arbitrarily small, see below.}

While our assertion bears a general nature, it is convenient to explain its origin using a simple model suggested in \cite{7}:
  scalar quantum electrodynamics supporting the Abrikosov-Nielsen-Olesen strings (Abrikosov strings in the
  nonrelativistic case) amended by a real scalar spin-1 field.  The Lagrangian will explicitly break the Lorentz boost part of
  the Poincar\'e group, but this is irrelevant for the phenomenon under consideration. We will focus on the spontaneous breaking of the O(3) rotational part.

The model is described by an effective 
Lagrangian
\beq
{\cal L} = {\cal L}_{\rm v} +{\cal L}_\chi
\eeq
where
 \beq
{\cal L}_{\rm v} = -\frac{1}{4e^2}F_{\mu\nu}^2 + \left| {\mathcal D}^\mu\phi\right|^2
  -V(\phi )\, ,\qquad V= \lambda \left(|\phi |^2 -v^2
\right)^2\,,
  \label{tpi16}
\eeq
and
\beqn
{\cal L}_\chi &=& \partial_\mu \chi^i \, \partial^\mu \chi^i -\varepsilon (\partial_i\chi^i)^2 - U(\chi, \phi)\,,
\qquad i=1,2,3\,,
\label{14}\\[2mm]
U &=&  \gamma\left[\left(-\mu^2 +|\phi |^2
\right)\chi^i \chi^i + \beta \left( \chi^i \chi^i\right)^2\right],
\label{15p}
\eeqn
with self-evident definitions of the fields involved, the covariant derivative, and the kinetic and potential terms.
One can consider both, relativistic and nonrelativistic kinetic terms. For definiteness we will focus on the first choice.
The parameter $\varepsilon$ will be treated as adjustable, 
while all other constants $e$, $\lambda$, $\beta$, $\mu$, and $v$ can be chosen at will, with some mild constraints
(e.g. $v>\mu$) discussed in \cite{7}. The model design is conceptually similar to that
of the superconducting cosmic strings\,\footnote{The only difference is the ``second" scalar field:
in \cite{7} it is   real and three-component while in \cite{Wi} it is a complex field similar to $\phi$.} in \cite{Wi}.

In the vacuum $\phi$ develops an expectation value $\phi_{\rm vac}= v$, while $\chi^i$ does {\em not} condense.
The global symmetries of the vacuum state are four translations and three O(3) rotations, in addition to the U(1) gauge symmetry which is implemented in the Higgs mode because the complex field $\phi$ is condensed.

If we disregard the $\chi^i$ fields, the model under consideration is the Abelian Higgs model which supports conventional flux tubes (strings), see  e.g. \cite{8}. They are topologically stable because of nontrivial windings of the $\phi$ field. 
The minimal flux tube is associated with a single winding.

In the core of such a tube the $\phi$ field tends to zero with necessity. Switching on the  $\chi^i$ fields we observe  that this implies the $\chi$ field  destabilization \cite{Wi} in the core (as follows from Eq.~(\ref{15p})). 
Hence, inside the core the $\chi$ field no longer vanishes. 

Our ``two-component" flux tube (string) spontaneously breaks two translational symmetries, in the perpendicular $x,y$ plane, and O(3) rotations. The latter are spontaneously broken by the string orientation along the $z$ axis (more exactly, O(3)$\to $O(2)), and by the orientation of the spin field $\chi^i$ inside the core of the flux tube.

In order to establish implications of this symmetry breaking let us start from the limit $\varepsilon \to 0$. In this limit, if we forget for a short while about the $\chi^i$ fields, the breaking O(3)$\to $O(2) produces no extra Goldstone modes
on the string world sheet. We have only two translational modes, because the order parameter is scalar, and the action of the broken rotational generators reduces to that of translational generators \cite{3}.

However, due to the fact that $\chi\neq 0$ in the core, the presence of the $\chi$ component in the string solution
leads to the occurrence of two extra moduli on the string world sheet. Indeed, in the limit $\varepsilon \to 0$
the rotational O(3) symmetry is enhanced \cite{6}: as a matter of fact,  one adds  {\em independent} O(3)
rotations of the spin field $\chi^i$ to the coordinate spacial rotations. 

The spin symmetry
is  (spontaneously)  broken down to O(2) on the string solution.
This latter breaking gives rise to the two-dimensional O(3) sigma model\footnote{Equivalent to the CP(1) model.} on the string world sheet (whose target space is
O(3)/O(2)). As a result, in addition to two translational moduli we have two orientational moduli in the
case at hand. Numerical calculations supporting the above assertion of the spontaneous breaking O(3)$\to$O(2)
in the string core will be presented elsewhere \cite{msy}.

Now, let us leave the limit $\varepsilon = 0$, and see what happens at $\varepsilon\neq 0$. If 
$\varepsilon$ small, to the leading order in this parameter, we can determine the effective world-sheet action using the
solution obtained at $\varepsilon = 0$. Note that at $\varepsilon\neq 0$ three Lorentz boosts
are explicitly broken in (\ref{15p}). In addition, two distinct O(3) rotations mentioned above become entangled: 
O(3)$\times$O(3) is no longer the exact symmetry of the model, but, rather, an approximate symmetry.

 Implications  on the string world sheet will ensue. Upon reflection it is easy to see that the
 term $\varepsilon (\partial_i\chi^i)^2$ lifts the degeneracy on the target space
 of the world-sheet O(3) sigma model. The low-energy effective action on the string world sheet is
 \beqn
 S&= &\int dt\,dz\left(  {\cal L}_{{\rm O}(3)} + {\cal L}_{x_\perp} \right),
  \nonumber\\[2mm]
{\cal L}_{{\rm O}(3)} &=& \left\{ \frac{1}{2g^2} \left[\left(\partial_a S^i\right)^2  + \varepsilon \left(\partial_zS^3\right)^2\right]
 \right\}
 -
 M^2 \left(1- (S^3)^2\right),
   \label{five}\\[2mm]
{\cal L}_{x_\perp}&=& \frac{T}{2} \left(\partial_a \vec x_\perp\right)^2 - \tilde{M}^2 \left(S^3\right)^2 \left(\partial_z \vec{x}_\perp\right)^2
 ,
 \label{six}
 \eeqn
where $\vec x_\perp =\{ x(t,z),\, y(t,z)\}$ are the translational moduli fields,
three orientational (quasi)moduli  fields $S^i(t,z)$ are constrained ($i=1,2,3$),
\beq
S^i\,S^i =1\,,
\eeq
$a=t,z$, are the string world-sheet coordinates, and $T$ is the string tension. The constants  $g^2$, $M^2$,  and $\tilde{M}^2$ are
\beq
g^2 \sim \beta\gamma\,,\qquad M^2 \sim \tilde{M}^2 \sim \varepsilon\,{\mu^2}/{\beta}\,,
\eeq
assuming $\mu^2\sim v^2$.
If $\varepsilon\to 0$ (i.e. $M^2=\tilde{M}^2 =0$)
we recover the standard O(3) sigma model, with the target space O(3)/O(2) and two moduli fields
(gapless excitations). With nonvanishing but small $\varepsilon $ the gapless rotational excitations become
quasigapless\,\footnote{We assume that $M^2\ll T$. At weak coupling in the bulk $\gamma \ll 1$ and, hence, $g^2\ll 1$. } (note that $M^2\sim \varepsilon$). The two-dimensional Lorentz boost 
is no longer a symmetry, since (as was mentioned above), the Lorentz boosts are explicitly broken by the
$\varepsilon (\partial_i\chi^i)^2$ term in four dimensions, see (\ref{15p}).

In high-energy physics $M^2$ is referred to as the twisted mass \cite{twisted}. In condensed matter the $\varepsilon =0$
limit of ${\cal L}_{{\rm O}(3)}$ is known as the Heisenberg antiferromagnet model. Then the last term in (\ref{five}) can be interpreted as
an external magnetic field of a special form giving rise to an isotropy term (e.g. \cite{A} and discussion therein). 

The impact of the mass term in (\ref{five}) depends on the sign of $M^2$ (inherited from $\varepsilon$).
If $M^2$ is positive the ground state of the theory -- the vacuum --
is achieved at $S^3=\pm 1$, i.e. the spin vector in the flux tube core is aligned with the tube axis (the so-called
easy axis). If $M^2$ is negative,
the ground state is achieved at $S^3=0$, i.e. the spin vector is perpendicular to the axis \cite{A} (the so-called easy plane).
Then the vacuum manifold is developed ($S_1$), the string becomes axially asymmetric, 
and instead of two quasigapless excitations we have one quasigapless and one classically gapless.\footnote{
It can acquire a mass gap, though, through the Berezinskii-Kosterlitz-Thouless phase transition.}

So far we discussed the theory on the string world sheet in the classical approximation.
It is well-known \cite{BP} that in the $\varepsilon\to 0$ limit the model (\ref{five}) is asymptotically free and develops
a dynamical mass gap $\Lambda$ as a result of infrared dynamics. The value of this mass gap is determined by the value of $g^2$ and can be made arbitrarily small with the appropriate choice of parameters.
If $\Lambda$ is smaller than $M$ the discussion above remains valid. Otherwise,
the mass gap is determined by $\Lambda$.  

\vspace{2mm}

We are grateful to Alex Kamenev, W. Vinci, and G. Volovik for inspirational  discussions.
The work of M.S. is supported in part by DOE grant DE-FG02- 94ER-40823. 
The work of A.Y. is  supported 
by  FTPI, University of Minnesota, 
by RFBR Grant No. 13-02-00042a 
and by Russian State Grant for 
Scientific Schools RSGSS-657512010.2.

\end{document}